\documentclass[12pt]{article}
\usepackage{oldgerm}
\usepackage{euler}
\usepackage{graphics}
\usepackage{bm}
\usepackage{graphicx}
\usepackage{amsmath}
\usepackage{amssymb}
\usepackage{amstext}
\usepackage{amscd}
\usepackage{amsfonts}
\usepackage{appendix}
\usepackage{color}
\DeclareMathAlphabet{\EuFrak}{U}{euf}{m}{n}
\DeclareMathAlphabet{\EuScript}{U}{eus}{m}{n}

\newcommand{\nd}{\noindent}

\newcommand{\be}{\begin{equation}}
\newcommand{\ee}{\end{equation}}
\newcommand{\ben}{\begin{eqnarray}}
\newcommand{\een}{\end{eqnarray}}

\title{{\bf q-Gamow States for intermediate energies}}

\author{{A. Plastino$^1$, M. C. Rocca$^1$ and D. J. Zamora$^1$}\\
\small{$^1$ Department of Physics,
La Plata National University}\\ \small{ and
Argentina's National Research Council}\\
\small{(IFLP-CCT-CONICET)-C. C. 727, 1900 La Plata - Argentina}}

\date{\today}

\begin{document}

\maketitle

\begin{abstract}

In a recent paper [Nuc. Phys. A {\bf 948}, (2016) 19] we have
demonstrated the possible existence of Tsallis' q-Gamow states. Now,
accelerators' experimental evidence for Tsallis' distributions has
been ascertained only at very high energies. Here, instead, we
develop a different set of q-Gamow states for which the associated
q-Breit-Wigner distribution could easily be found at intermediate
energies, for which accelerators are available at many locations. In
this context, it should be strongly emphasized [Physica A {\bf 388}
(2009) 601] that, empirically, one never exactly and unambiguously  ''detects'' pure
Gaussians, but rather q-Gaussians.   A prediction is made via Eq. \ref{predict}.

\nd {\bf Keywords:} Gamow States, q-Gamow States, q-Gaussians.

\end{abstract}

\newpage

\renewcommand{\theequation}{\arabic{section}.\arabic{equation}}

\section{Introduction}

Empirical analysis abundantly shows  that power-law behavior in the
(observed) distribution of some quantity is rather common in nature
\cite{boccara}. It has been shown the a  reason
 for this high frequency is  detector-normalization \cite{vp}. Since very many systems
are statistically described by power-law probability distributions
\cite{cero}, this is a subject that deserves attention. In
particular, q-Gaussian behavior is frequently found  in different
scenarios. Reference \cite{vp} explains why. This happens in
experimental scenarios for which data are gathered using a set-up
that performs a normalization-preprocessing. \cite{vp} finds that in
such settings  the value of the associated parameter $q$ can be
deduced from  the normalization technique that characterizes the
empirical device. Of course, by a q-exponential we mean the function

\be e_q(x)= [1 + (1-q)x]^{1/(1-q)}, \ee that tends to the ordinary
exponential when $q$ approaches unity \cite{tsallis}.

It was shown in \cite{tp1,tp2,tp3,tq1} that resonances, i.e.
Gamow-states \cite{tp4,tq1p},  can be seen  as (Sebastiao e Silva's)
Ultradistributions  \cite{tp5,tp6,tp7}. Their treatment needs
appealing to Rigged Hilbert Space \cite{tp8,tp9,tp10,tq2}. It was
demonstrated in \cite{npa} that associated resonances, called
q-Gamow states, constitute a useful generalization of the GS concept
adapted to Tsallis' q-statistics \cite{tsallis}.

 Indeed, one can find a large
number of high energy experiments amenable to  interpretation via
Tsallis' q-statistics \cite{tsallis}, specifically,  LHC experiments
in what concerns distributions connected to stationary states.
Tsallis' q-statistics  adequately describes the transverse momentum
distributions of variegated hadrons. All four  LHC experiments
generated papers using such distributions that can be adequately
fitted  employing  the q-exponential function. The pertinent q-value
is of the order of 1.15, clearly distinct  from the unity value of
Gibbs-Boltzmann's statistics. This evidences that  stationary states
before hadronization can not be thermal equilibrium-ones
\cite{tsallis}. Measurement of the $p_T$ distribution over a
logarithmic range of fourteen decades demonstrates  that $q=1.15$
fits nicely  the data over this large  range \cite{tp11,phenix}).

 These findings  motivate one to study  complex energy states
  related to      the  q-exponential distributions (namely,  q-Gamow
states), that are not solutions of
Schroedinger's equation but of its non linear, q-version counterpart, advanced Nobre, Rego-Monteiro, and
 Tsallis in \cite{nobre} (see also  \cite{tp15}). We recommend, for a review on ordinary Gamow
states, reference \cite{tw1}.

Remark that accelerators' experimental evidence for Tsallis' distributions has
been ascertained only at very high energies. This motivates us to inquire about other kind of q-Gamow states
 for which the associated
q-Breit-Wigner distribution could easily be found at intermediate
energies, for which accelerators are available at many locations.
 Such is then the goal we seek to achieve here. WE base our
 considerations on the fact that, empirically, one does not often detect pure   Gaussians, but rather q-Gaussians
\cite{vp}. This suggests looking for
special q-Gamow states in the
q-neighborhood of $q=1$.

\setcounter{equation}{0}

\section{New q-Gamow States to be introduced here}

We obtain a new kind of Gamow state for  $q$ close to unity, via perturbation
theory around such $q-$value, of the states studied in  \cite{npa},
keeping up to first order terms. Accordingly,
\begin{equation}
\label{ep2.1} \left[1+\frac {i(1-q)px}
{\hbar\sqrt{2(q+1)}}\right]^{\frac {2} {1-q}} \simeq
\left[1-(q-1)\left(\frac {ipx} {4\hbar}+ \frac {p^2x^2}
{4\hbar^2}\right)\right] e^{\frac {ipx} {\hbar}},
\end{equation}
and the ensuing new q-Gamow state becomes
\[|\psi_{qG}>=\int\limits_{-\infty}^{\infty}
\left\{{\cal H}[\Im(p)]{\cal H}(x)-{\cal H}[-\Im(p)]{\cal H}(-x)\right\}
\otimes\]
\begin{equation}
\label{ep2.2} \left[1-(q-1)\left(\frac {ipx} {4\hbar}+ \frac
{p^2x^2} {4\hbar^2}\right)\right] e^{\frac {ipx} {\hbar}}|x>dx,
\end{equation}
or
\[\psi_{qG}(x)=
\left\{{\cal H}[\Im(p)]{\cal H}(x)-{\cal H}[-\Im(p)]{\cal H}(-x)\right\}
\otimes\]
\begin{equation}
\label{ep2.3} \left[1-(q-1)\left(\frac {ipx} {4\hbar}+ \frac
{p^2x^2} {4\hbar^2}\right)\right] e^{\frac {ipx} {\hbar}}.
\end{equation}
The  norm of a  q-Gamow state is now
\[<\psi_{qG}|\psi_{qG}>=\]
\[\int\limits_0^{\infty}
{\cal H}[\Im(p)]
\left[1-(q-1)\left(\frac {ipx} {4\hbar}+
\frac {p^2x^2} {4\hbar^2}\right)\right]
e^{\frac {ipx} {\hbar}}\otimes\]
\[\left[1+(q-1)\left(\frac {ip^{\ast}x} {4\hbar}-
\frac {p^{\ast 2}x^2} {4\hbar^2}\right)\right]
e^{-\frac {ip^{\ast}x} {\hbar}}
\;dx \]
\[+\int\limits_{-\infty}^0 {\cal H}[-\Im(p)]
\left[1-(q-1)\left(\frac {ipx} {4\hbar}+
\frac {p^2x^2} {4\hbar^2}\right)\right]
e^{\frac {ipx} {\hbar}}\otimes\]
\begin{equation}
\label{ep2.4} \left[1+(q-1)\left(\frac {ip^{\ast}x} {4\hbar}- \frac
{p^{\ast 2}x^2} {4\hbar^2}\right)\right] e^{-\frac {ip^{\ast}x}
{\hbar}} \;dx,
\end{equation}
or equivalently,
\[<\psi_{qG}|\psi_{qG}>=\]
\[\int\limits_0^{\infty}
{\cal H}[\Im(p)]
\left[1-(q-1)\left(\frac {ipx} {4\hbar}+
\frac {p^2x^2} {4\hbar^2}\right)\right]
e^{\frac {ipx} {\hbar}}\otimes\]
\[\left[1+(q-1)\left(\frac {ip^{\ast}x} {4\hbar}-
\frac {p^{\ast 2}x^2} {4\hbar^2}\right)\right]
e^{-\frac {ip^{\ast}x} {\hbar}}
\;dx \]
\[+\int\limits_0^{\infty} {\cal H}[-\Im(p)]
\left[1+(q-1)\left(\frac {ipx} {4\hbar}-
\frac {p^2x^2} {4\hbar^2}\right)\right]
e^{-\frac {ipx} {\hbar}}\otimes\]
\begin{equation}
\label{ep2.5} \left[1-(q-1)\left(\frac {ip^{\ast}x} {4\hbar}+ \frac
{p^{\ast 2}x^2} {4\hbar^2}\right)\right] e^{\frac {ip^{\ast}x}
{\hbar}} \;dx.
\end{equation}
After a little algebra, (\ref{ep2.5}) becomes
\[<\psi_{qG}|\psi_{qG}>=\]
\[\int\limits_0^{\infty}
{\cal H}[\Im(p)]
\left\{1+(q-1)\left[\frac {i(p^{\ast}-p)x} {4\hbar}-
\frac {(p^2+p^{\ast 2})x^2} {4\hbar^2}\right]
\right\}e^{\frac {i(p-p^{\ast})x} {\hbar}}\;dx+\]
\begin{equation}
\label{ep2.6} \int\limits_0^{\infty} {\cal H}[-\Im(p)]
\left\{1+(q-1)\left[\frac {i(p-p^{\ast})x} {4\hbar}- \frac
{(p^2+p^{\ast 2})x^2} {4\hbar^2}\right] \right\}e^{\frac
{i(p^{\ast}-p)x} {\hbar}}\;dx.
\end{equation}
Integration is straightforward:
\begin{equation}
\label{ep2.7} <\psi_{qG}|\psi_{qG}>= \frac {\hbar} {2|\Im(p)|}
\left\{1+\frac {(q-1)} {4}\left[ 1+\frac {\Im(p)^2-\Re(p)^2}
{\Im(p)^2}\right]\right\}.
\end{equation}
Thus, squaring the norm we find
\begin{equation}
\label{ep2.8} A^2(q,p)= \frac {\hbar} {2|\Im(p)|} \left\{1+\frac
{(q-1)} {4}\left[ 1+\frac {\Im(p)^2-\Re(p)^2}
{\Im(p)^2}\right]\right\},
\end{equation}
and then one has
\begin{equation}
\label{ep2.9} A(q,p)= \sqrt{\frac {\hbar} {2|\Im(p)|} \left\{1+\frac
{(q-1)} {4}\left[ 1+\frac {\Im(p)^2-\Re(p)^2}
{\Im(p)^2}\right]\right\}}.
\end{equation}
Note that:
\begin{equation}
\label{ep2.10} \lim_{q\rightarrow 1}A(q,p)=\sqrt{\frac {\hbar}
{2|\Im(p)|}}.
\end{equation}
We see that  (\ref{ep2.10}) and (\ref{eqa.3}) agree The normalized
q-Gamow state is now

\begin{equation} \label{ep2.11}
|\phi_{qG}>=\frac {|\psi_{qG}>} {A(q,p)},
\end{equation}
that can be recast in the fashion

\[|\phi_{qG}>=\int\limits_{-\infty}^{\infty}
\left\{{\cal H}[\Im(p)]{\cal H}(x)-{\cal H}[-\Im(p)]{\cal H}(-x)\right\}
\sqrt{\frac {2|\Im(p)|} {\hbar}}
\otimes\]
\begin{equation}
\label{ep2.12} \left[1-(q-1)\left( \frac {1} {8}+\frac
{\Im(p)^2-\Re(p)^2} {8\Im(p)^2}+ \frac {ipx} {4\hbar}+ \frac
{p^2x^2} {4\hbar^2}\right)\right] e^{\frac {ipx} {\hbar}}|dx>,
\end{equation}
and thus

\[\phi_{qG}(x)= \left\{{\cal H}[\Im(p)]{\cal H}(x)-{\cal
H}[-\Im(p)]{\cal H}(-x)\right\} \sqrt{\frac {2|\Im(p)|} {\hbar}}
\otimes\]
\begin{equation}
\label{ep2.13} \left[1-(q-1)\left( \frac {1} {8}+\frac
{\Im(p)^2-\Re(p)^2} {8\Im(p)^2}+ \frac {ipx} {4\hbar}+ \frac
{p^2x^2} {4\hbar^2}\right)\right] e^{\frac {ipx} {\hbar}}.
\end{equation}
\vskip 3mm

We show now that the new q-Gamow states we are speaking about here
satisfy the nonlinear q-Schroedinger equation of reference
\cite{tp15}. Consider the function $f(x)$ defined by

\begin{equation}
\label{ep2.14} f(x)=\sqrt{\frac {2|\Im(p)|} {\hbar}}
\left[1-(q-1)\left( \frac {1} {8}+\frac {\Im(p)^2-\Re(p)^2}
{8\Im(p)^2}+ \frac {ipx} {4\hbar}+ \frac {p^2x^2}
{4\hbar^2}\right)\right] e^{\frac {ipx} {\hbar}}.
\end{equation}
We wish to show that  $f(x)$ satisfies

\begin{equation}
\label{ep2.15}
H\left[\frac {f(x)} {f(0)}\right]=\frac {p^2} {2m}
\left[\frac {f(x)} {f(0)}\right]^q,
\end{equation}
or

\begin{equation}
\label{ep2.16} Hf(x)=\frac {p^2} {2m} [f(0)]^{1-q} [f(x)]^q,
\end{equation}
taking into account that

\begin{equation}
\label{ep2.17} [f(0)]^{1-q}=1-(q-1) \ln\left[\sqrt{\frac {2|\Im(p)|}
{\hbar}}\right],
\end{equation}
and

\[[f(x)]^q=\sqrt{\frac {2|\Im(p)|} {\hbar}}
e^{\frac {ipx} {\hbar}}\otimes\]
\begin{equation}
\label{ep2.18} \left\{1+(q-1)\left[ \ln\left(\sqrt{\frac {2|\Im(p)|}
{\hbar}}\right)- \frac {1} {8}-\frac {\Im(p)^2-\Re(p)^2} {8\Im(p)^2}
+\frac {3ipx} {4\hbar}-\frac {p^2x^2} {4\hbar^2} \right]\right\}.
\end{equation}
Accordingly, we find

\[[f(0)]^{1-q}[f(x)]^q=\sqrt{\frac {2|\Im(p)|} {\hbar}}
e^{\frac {ipx} {\hbar}}\otimes\]
\[\left\{1-(q-1)\left[
\frac {1} {8}+\frac {\Im(p)^2-\Re(p)^2} {8\Im(p)^2}
-\frac {3ipx} {4\hbar}+\frac {p^2x^2} {4\hbar^2}
\right]\right\}=\]
\begin{equation}
\label{ep2.19} -\frac {1} {p^2}\frac {d^2f(x)} {dx^2}= \frac {2m}
{p^2}Hf(x).
\end{equation}
Minding  (\ref{ep2.19}) we see that  $f(x)$ satisfies (\ref{ep2.15})
and, consequently,  the  q-Gamow states also verify it. \vskip 3mm

We pass now to compute the mean value of the  energy corresponding
to a  q-Gamow state. We begin with

\[H|\phi_{qG}>=
\frac {p^2} {2m}
\sqrt{\frac {2|\Im(p)|} {\hbar}}
\otimes\]
\[\int\limits_{-\infty}^{\infty}
\left\{{\cal H}[\Im(p)]{\cal H}(x)-{\cal H}[-\Im(p)]{\cal H}(-x)\right\}
\otimes\]
\begin{equation}
\label{ep2.20} \left\{1-(q-1)\left[ \frac {1} {8}+\frac
{\Im(p)^2-\Re(p)^2} {8\Im(p)^2} -\frac {3ipx} {4\hbar}+\frac
{p^2x^2} {4\hbar^2} \right]\right\} e^{\frac {ipx} {\hbar}}|x>dx,
\end{equation}
and thus

\[<\phi_{qG}(H|\phi_{qG}>)=
\frac {p^2} {2m}
\frac {2|\Im(p)|} {\hbar}
\otimes\]
\[\int\limits_{-\infty}^{\infty}
\left\{{\cal H}[\Im(p)]{\cal H}(x)-{\cal H}[-\Im(p)]{\cal H}(-x)\right\}
\otimes\]
\[\left\{1-(q-1)\left[
\frac {1} {8}+\frac {\Im(p)^2-\Re(p)^2} {8\Im(p)^2}
-\frac {ip^{\ast}x} {4\hbar}+\frac {p^{\ast 2}x^2} {4\hbar^2}
\right]\right\}\otimes\]
\begin{equation}
\label{ep2.21} \left\{1-(q-1)\left[ \frac {1} {8}+\frac
{\Im(p)^2-\Re(p)^2} {8\Im(p)^2} -\frac {3ipx} {4\hbar}+\frac
{p^2x^2} {4\hbar^2} \right]\right\} e^{\frac {i(p-p^{\ast})x}
{\hbar}}dx.
\end{equation}
The preceding  equation   can be recast as
\[<\phi_{qG}(H|\phi_{qG}>)=
\frac {p^2} {2m}
\frac {2|\Im(p)|} {\hbar}
\otimes\]
\[\left\{\int\limits_0^{\infty}
{\cal H}[\Im(p)]
\left\{1-(q-1)\left[
\frac {1} {4}+\frac {\Im(p)^2-\Re(p)^2} {4\Im(p)^2}\right.\right.\right.\]
\[\left.\left.-\frac {i(3p+p^{\ast})x} {4\hbar}+
\frac {(p^2+p^{\ast 2})x^2}
{4\hbar^2}
\right]\right\}e^{\frac {i(p-p^{\ast})x} {\hbar}}dx\]
\[\int\limits_0^{\infty}
{\cal H}[-\Im(p)]\left\{1-(q-1)\left[
\frac {1} {4}+\frac {\Im(p)^2-\Re(p)^2} {4\Im(p)^2}\right.\right.\]
\begin{equation}
\label{ep2.22} \left.\left.\left.+\frac {i(3p+p^{\ast})x} {4\hbar}+
\frac {(p^2+p^{\ast 2})x^2} {4\hbar^2} \right]\right\} e^{\frac
{i(p^{\ast}-p)x} {\hbar}}dx\right\}.
\end{equation}
Evaluating the integrals in  (\ref{ep2.22}) we encounter
\begin{equation}
\label{ep2.23} <\phi_{qG}|(H|\phi_{qG}>)=\frac {p^2} {2m}
\left\{1-(q-1)\left[\frac {1} {4}- \frac {i(3p+p^{\ast})}
{8|\Im(p)|}Sgn[\Im(p)]\right]\right\}.
\end{equation}
Analogously, we reach

\begin{equation}
\label{ep2.24} (<\phi_{qG}|H)|\phi_{qG}>=\frac {p^{\ast 2}} {2m}
\left\{1-(q-1)\left[\frac {1} {4}- \frac {i(3p^{\ast}+p)}
{8|\Im(p)|}Sgn[\Im(p)]\right]\right\}.
\end{equation}
Thus, according to  \cite{npa}, we obtain for the mean energy value
\begin{equation}
\label{ep2.25} <H>_q=\frac {1} {2} \left[<\phi_{qG}|(H|\phi_{qG}>)+
(<\phi_{qG}|H)|\phi_{qG}>\right].
\end{equation}
Additionally, we have
\begin{equation}
\label{ep2.26}
\lim_{q\rightarrow 1}<H>_q=\frac {\Re(p^2)} {2m}=<H>.
\end{equation}

\section{Prediction: q-Breit-Wigner distribution}

 We compute now the pertinent new q-Breit-Wigner
distribution. We begin with

\[<\phi|\phi_{Gq}>=\frac {1} {\hbar}
\sqrt{\frac {|\Im(p)|} {\pi}}
\left\{
\int\limits_{-\infty}^{\infty}
\left\{{\cal H}[\Im(p)]{\cal H}(x)-{\cal H}[-\Im(p)]{\cal H}(-x)\right\}
\right.\]
\begin{equation}
\label{ep2.27} \left[1-(q-1)\left( \frac {1} {8}+\frac
{\Im(p)^2-\Re(p)^2} {8\Im(p)^2}+ \frac {ipx} {4\hbar}+ \frac
{p^2x^2} {4\hbar^2}\right)\right] e^{\frac {i(p-k)x} {\hbar}}dx.
\end{equation}
After evaluation the integrals in  (\ref{ep2.27}) we have
\[<\phi|\phi_{Gq}>=\frac {1} {i(k-p)}
\sqrt{\frac {|\Im(p)|} {\pi}}\otimes\]
\begin{equation}
\label{ep2.28} \left[1-(q-1)\left( \frac {1} {8}+\frac
{\Im(p)^2-\Re(p)^2} {8\Im(p)^2}+ \frac {p} {4(k-p)}- \frac {p^2}
{2(k-p)^2}\right)\right].
\end{equation}
Thus, the q-Breit-Wigner relation is
\[|<\phi|\phi_{Gq}>|^2=
\frac
{|\Im(p)|} {\pi\left\{[\Re(p)-k]^2+\Im(p)^2\right\}}\]

\begin{equation}
\label{ep2.29} \left[1-(q-1)\left( \frac {1} {4}+\frac
{\Im(p)^2-\Re(p)^2} {4\Im(p)^2}+ \frac {p+p^{\ast}} {4(k-p)}- \frac
{p^2+p^{\ast 2}} {2(k-p)^2}\right) \right].
\end{equation}

The factor $X$

\be X= (q-1)\left( \frac {1} {4}+\frac{\Im(p)^2-\Re(p)^2} {4\Im(p)^2}+ \frac {p+p^{\ast}} {4(k-p)}- \frac
  {p^2+p^{\ast 2}} {2(k-p)^2}\right),  \label{predict}\ee
  constitutes the signature of our new q-Gamow resonances and is, in
  principle, amenable of empirical verification.
   \vskip 3mm

Note that for $q\rightarrow 1$ one has
\begin{equation}
\label{eq2.30} \lim_{q\rightarrow 1}|<\phi|\phi_{Gq}>|^2=\frac
{|\Im(p)|} {\pi\left\{[\Re(p)-k]^2+\Im(p)^2\right\}}.
\end{equation}
in agreement with  (\ref{eqa.10}).

\setcounter{equation}{0}

\section{Conclusions}

It is the essence to point out that, as discussed in \cite{vp},
empirically one often obtains  q-Gaussians rather than pure Gaussians, with
$q$ very close to one. Accordingly,  for a q-region in the immediate neighborhood of
$q=1$ we have here studied the main properties of the associated
q-Gamow states, that are solutions to the NRT-nonlinear, q-generalization of
Schroedinger's equation  \cite{nobre,tp15}.       \vskip 3mm

We have computed their norm, the mean
energy value, and the concomitant   q-Breit-Wigner distributions.
 In all instances, results tend to the customary ones when
 the all important $q$-parameter of Tsallis' obeys $q\rightarrow
 1$.                                                                                                                                                         Accordingly, in this effort we  introduced new intermediate energy q-Gamow states,

   \vskip 3mm

Our main result is that our q-Breit-Wigner probability distribution will differ from the usual one
 according to the factor \ref{predict}, which might be checked out, after careful error's analysis,
  in extant accelerator data,
  thus proving the existence of the new q-Gamow states we are advancing here.

{\bf Acknowledgment:}  The authors thank Conicet grant  PIP029/12.

\newpage

\renewcommand{\thesection}{\Alph{section}}

\renewcommand{\theequation}{\Alph{section}.\arabic{equation}}

\setcounter{section}{1}

\section*{Appendix}

\subsection*{Review on Gamow States}

This appendix summarizes results from  \cite{npa}.

A  Gamow state, at large distances from the scattering center has
the form

\begin{equation}
\label{eqa.1} |\psi_G>=\int\limits_{-\infty}^{\infty} \left\{{\cal
H}[\Im(p)]{\cal H}(x)-{\cal H}[-\Im(p)]{\cal H}(-x)\right\} e^{\frac
{ipx} {\hbar}}|x>\;dx.
\end{equation}
The square of the norm reads
\begin{equation}
\label{eqa.2} <\psi_G|\psi_G>=\int\limits_0^{\infty} {\cal
H}[\Im(p)]e^{\frac {i(p-p^{\ast})x} {\hbar}}\;dx -
\int\limits_{-\infty}^0 {\cal H}[-\Im(p)]e^{\frac {i(p-p^{\ast})x}
{\hbar}}\;dx.
\end{equation}
These  integrals can be easily evaluated. One finds
\begin{equation}
\label{eqa.3} <\psi_G|\psi_G>=\left\{{\cal H}[\Im(p)]- {\cal
H}[-\Im(p)]\right\}\frac {\hbar} {i(p^{\ast}-p)}= \frac {\hbar}
{2|\Im(p)|}.
\end{equation}
Accordingly, the normalized Gamow-state  $\phi_G$ becomes
\cite{tq1p}
\begin{equation}
\label{eqa.4} |\phi_G>=\sqrt{\frac {2|\Im(p)|} {\hbar}}|\psi_G>.
\end{equation}
Additionally,
\begin{equation}
\label{eqa.5}
 <\phi_G|(H|\phi_G>)=\frac {p^2} {2m},
\end{equation}
\begin{equation}
\label{eqa.6}
(<\phi_G|H)|\phi_G>=\frac {p^{\ast 2}} {2m},
\end{equation}
The mean energy is
\begin{equation}
\label{eqa.7}
<H>=\frac {1}
{2}\left[<\phi_G|(H|\phi_G>)+(<\phi_G|H)|\phi_G>\right]= \frac
{p^2+p^{\ast 2}} {4m}= \frac {\Re(p^2)} {2m}.
\end{equation}
In order to obtain de probability distribution associated to a Gamow
state we start by the looking at the scalar product between this
state and a free one
\begin{equation}
\label{eqa.8}
<\phi|\phi_G>=\frac {1} {\hbar}\sqrt{\frac
{|\Im(p)|} {\pi}} \left\{\int\limits_0^{\infty} {\cal
H}[\Im(p)]e^{\frac {i(p-k)x} {\hbar}}\;dx- \int\limits_{-\infty}^0
{\cal H}[-\Im(p)]e^{\frac {i(p-k)x} {\hbar}}\;dx\right\}.
\end{equation}
Thus,
\begin{equation}
\label{eqa.9}
<\phi|\phi_G>=\frac {i\sqrt{\frac {|\Im(p)|} {\pi}}}
{p-k}
\end{equation}
The ensuing probability distribution is the  Breit-Wigner one
\cite{tq1p}
\begin{equation}
\label{eqa.10} |<\phi|\phi_G>|^2=\frac {|\Im(p)|}
{\pi\left\{[\Re(p)-k]^2+\Im(p)^2\right\}}.
\end{equation}

\end{document}